\documentclass[manuscript]{acmart}
\makeatletter
\renewcommand\@formatdoi[1]{\ignorespaces}
\makeatother

\AtBeginDocument{%
  \providecommand\BibTeX{{%
    \normalfont B\kern-0.5em{\scshape i\kern-0.25em b}\kern-0.8em\TeX}}}

\copyrightyear{2023}
\setcopyright{acmcopyright}
\acmYear{2023}
\acmDOI{XXXXXXX.XXXXXXX}
\acmConference[The 1st International Workshop on Explainable AI for the Arts]{}{ACM C\&C Conference}{Online}
\acmBooktitle{the 1st International Workshop on Explainable AI for the Arts (XAIxArts), ACM Creativity and Cognition (C\&C) 2023}
\acmPrice{15.00}
\acmISBN{978-1-4503-XXXX-X/18/06}

\begin{document}

\title[User-centric AIGC products]{User-centric AIGC products:  Explainable Artificial Intelligence and AIGC products}

\author{Hanjie Yu}
\authornotemark[1]
\email{yuhanjie.yhj@gmail.com}
\orcid{0009-0000-9085-5122}
\author{Yan Dong}
\authornote{Both authors contributed equally to this research.}
\email{m18643428545@163.com}
\orcid{0009-0008-5989-4201}

\affiliation{%
  \institution{Tsinghua University}
  \country{China}
}

\author{Qiong Wu}
\orcid{0000-0002-0304-5330}
\affiliation{%
  \institution{Tsinghua University}
  \country{China}
}

\renewcommand{\shortauthors}{Yan and Hanjie, et al.}

\begin{abstract}
Generative AI tools, such as ChatGPT and Midjourney, are transforming artistic creation as AI-art integration advances. However, Artificial Intelligence Generated Content (AIGC) tools face user experience challenges, necessitating a human-centric design approach. This paper offers a brief overview of research on explainable AI (XAI) and user experience, examining factors leading to suboptimal experiences with AIGC tools. Our proposed solution integrates interpretable AI methodologies into the input and adjustment feedback stages of AIGC products. We underscore XAI's potential to enhance the user experience for ordinary users and present a conceptual framework for improving AIGC user experience.
\end{abstract}

\ccsdesc[300]{Human-centered computing~HCI theory, concepts and models}

\keywords{Explainable Artificial Intelligence, Artificial Intelligence Generated Content, Product User Experience}



\maketitle

\section{Introduction}
AIGC products have been increasingly becoming an indispensable assistive creative toolkit for art and design. They provide designers with inspiration, visual style explorations, and accelerated workflow. However, AIGC faces challenges of transparency and interpretability, which impact user experience in ways including high learning costs, limited effectiveness and trust issues. The "black box" nature of the AIGC generation process makes it difficult for ordinary users (unfamiliar with artificial intelligence technology) to understand the specific reasons for the poor performance of artificial intelligence, hindering effective adjustment and utilization. Furthermore, numerous factors contribute to disparate AIGC results, yielding distinct responses for identical input. With inherent uncertainty and error rates in the generated results and widespread skepticism on AI ethics in public opinion, users often struggle to understand and trust AI.

XAI emerges as a solution to enhance user experiences in AIGC products. By integrating XAI into AIGC products, it helps users understand some algorithmic decision-making processes and principles to overcome challenges and improve user experience.Our mixed-methods research approach includes literature review, case studies, and user research. Key contributions are:
(1) An overview of the relationship between XAI and user experience;
(2) An analysis of causes for poor user experience with AIGC tools;
(3) A feasibility study on integrating interpretable AI approaches into AIGC user experience enhancement.

\section{Research Status}
\subsection{Dissonance between AIGC product usage and traditional user mental models}
Traditional product interaction design establishes mental models for users to predict system behavior and interact effectively. However, using AIGC products involves uncertainty and complexity \cite{yang2020re}, making feedback challenging to explain. AI systems may generate varying responses due to factors like lighting, noise, or text input variations, causing difficulties for users to comprehend these differences and transfer past experiences. People are prone to common cognitive biases (e.g., anchoring bias) after observing AI model behavior \cite{nourani2021anchoring}. These biases will exacerbate user frustration during re-operation.

This uncertainty also contradicts the principles of "Visibility of system status", "Consistency and standards", etc in traditional human-computer interaction \cite{nielsen2005ten}. By providing explanations for the AIGC generation process, such as local and contrastive explanations, we aim to alleviate this conflict and help users form mental models.

\subsection{XAI and User Experience}
Most XAI user experiences focus on model developers and managers use \cite{liao2021human}, typically using an algorithm-centric approach \cite{shin2021effects}, with little involvement from ordinary users and unable to help ordinary users understand AI products.

To promote AIGC products among ordinary users, XAI must adopt a human-centered approach. Miller \cite{miller2019explanation} identified characteristics of human-friendly explanations from a sociological perspective, while Amershi et al.\cite{amershi2019guidelines} proposed 18 usability guidelines for AI systems. However, these works mostly lack actionable guidance for product and usage improvements. Currently, there are no specific suggestions for product and usage improvements, such as which XAI technology might be integrated into which stage of the product. We will provide practice-based recommendations for potential subsequent improvements, combining XAI with the AIGC tool's usage process.

\section{Integrating XAI Technology into AIGC Product User Experience}
\begin{figure}[h]
  \centering
  \includegraphics[width=0.9\linewidth]{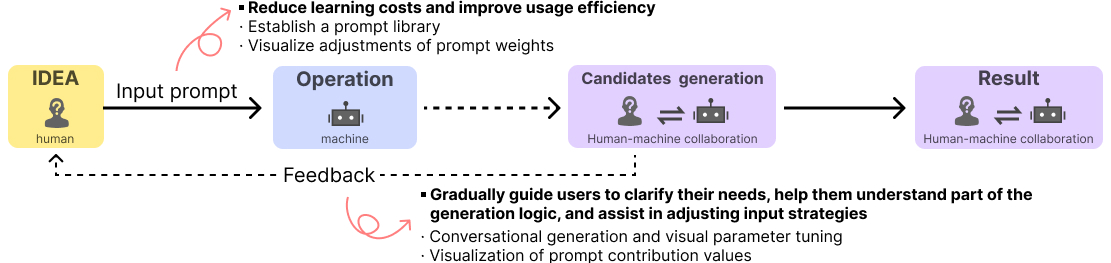}
  \caption{Simplified flow chart of AIGC product use.}
  \label{fig:fig1}
\end{figure}

AIGC products' workflow consists of three main parts, shown in Fig.~\ref{fig:fig1}. The conventional AIGC workflow is a unidirectional, iterative process involving user input prompts and machine-generated results. However, this process lacks a clear mapping between input and output, as well as feedback from the machine-generated. This paper proposes integrating XAI techniques into the AIGC workflow to enhance user experience during input and feedback stages.

\subsection{Optimization of user experience in the input stage: offering keyword libraries and prompt structures}
For novice users, inputting prompts may be challenging due to three factors: (1) Unclear instructions: AIGC tools may lack clear guidance, hindering users from providing effective input prompts. (2) Inaccurate generated content: AIGC tools might misinterpret input prompts, leading to content misaligned with user expectations. (3) Trial and error: Users often adjust input prompts multiple times to achieve satisfactory results. These issues stems from novice users' inability to map input words to output results. In traditional products, users anticipate outcomes using existing knowledge, while in AIGC, machines replace the production stage, making adjustments less predictable and controllable for users. A prompt library and weighted input prompts could help alleviate these issues.

One existing solution is to draw from the experience of creator communities regarding word combinations. Drawing from creator communities' experience, AIGC models can offer keyword libraries and prompt structures to facilitate basic effects and reduce trial-and-error costs. Libraries should encompass essential elements, including lighting, color tones, content, style, camera parameters and etc., and provide comparative style examples guided by keywords. Users can select keywords and fill into provided structures, while experienced users add self-descriptive keywords. Despite improved efficiency, prompt libraries may restrict flexibility and personalization. Therefore, a local explanation model to partially generalize the relationship between inputted keywords and outputs. Based on this, establishing a graphic user interface (resembles a color palette) enables users to assign varying weights to different parameters. 

\subsection{Optimization in the adjustment feedback stage: quantifying prompt contributions}
In traditional design methods, the primary challenge for designers is to address ill-defined problems (Double Dlamond,Google sprint,The five steps of design thinking by D.school) by repeatedly conceptualizing and iterating to understand what users truly want, and then presenting creative solutions.  In existing AIGC products, users are expected to define expectations clearly and translate them into machine-understandable descriptions. However, most users even struggle to identify their needs.  XAI technology can help AI guide users in clarifying their requirements through conversation, expressing expectations in a machine-understandable language.

Consequently, conversational generation is a promising approach that aligns with human thinking patterns and adjusts parts of the image with each conversation \cite{brooks2022instructpix2pix}, gradually aligning results with user requirements. However, current AIGC products face challenges in efficiently and precisely adjusting strategies or parameters when there is a significant gap between generated results and user expectations.Integrating sensitivity analysis and SHAP explainability methods into AIGC tools can improve user experience by elucidating generation logic and refining input strategies. By quantifying prompt word contributions and offering adjustable feature values, users can comprehend the significance and reasoning of each word, identify improvements, and boost efficiency. Future research can visualize parameter tuning results, establishing a more intuitive relationship between parameters and outputs, offering analysis and suggestions for improvement, and facilitating finer user adjustments.

\section{Conclusion}
AIGC is one of the fastest-growing fields in artificial intelligence. However, it currently operates in a technology-centric reality, and significant progress requires a focus on enhancing user experience. The XAI approaches mentioned in this paper for the input and adjustment stages aim to propose feasible directions for user experience improvement. These insights offer valuable guidance for researchers and practitioners in the application of XAI and the enhancement of AIGC user experiences.

\begin{acks}
This work is supported by the National Social Science Fund of China, Art Project (19BG127)
\end{acks}


\bibliographystyle{ACM-Reference-Format}
\bibliography{sample-base}

\appendix

\end{document}